\documentclass[ms]{copernicus}

\newcommand{\hii}{H\,\textsc{ii}}
\newcommand{\kms}{\ensuremath{\mathrm{km}\,\mathrm{s}^{-1}}}
\newcommand{\mr}[1]{\ensuremath{\mathrm{#1}}}
\newcommand{\pcmc}{\ensuremath{\mathrm{cm}^{-3}}}
\newcommand{\pcms}{\ensuremath{\mathrm{cm}^{-2}}}
\newcommand{\msun}{\ensuremath{\mathrm{M}_{\odot}}}
\newcommand{\msunperyr}{\ensuremath{\msun\,\mathrm{yr}^{-1}}}
\begin{document}

\nolinenumbers

\title{Effects of stellar evolution and ionizing radiation on the environments of massive stars}

\author[1]{Jonathan Mackey}
\author[1]{Norbert Langer}
\author[2]{Shazrene Mohamed}
\author[3]{Vasilii V.\ Gvaramadze}
\author[4]{Hilding R.\ Neilson}
\author[1]{Dominique M.-A.\ Meyer}

\affil[1]{Argelander-Institut f\"ur Astronomie, Auf dem H\"ugel 71, 53121 Bonn, Germany.}
\affil[2]{South African Astronomical Observatory, P.O.\ box 9, 7935 Observatory, South Africa.}
\affil[3]{Sternberg Astronomical Institute, Lomonosov Moscow State University, Universitetskij Pr.~13, Moscow 119992, Russia.}
\affil[4]{Department of Physics and Astronomy, East Tennessee State University, Box 70652, Johnson City, TN, 37614, USA}


\runningtitle{Environments of massive stars}

\runningauthor{J.\ Mackey \textit{et al.}}

\correspondence{Jonathan Mackey\\ (jmackey@astro.uni-bonn.de)}

\received{}
\pubdiscuss{} 
\revised{}
\accepted{}
\published{}


\firstpage{1}

\maketitle  

\begin{abstract}
We discuss two important effects for the astrospheres of runaway stars: the propagation of ionizing photons far beyond the astropause, and the rapid evolution of massive stars (and their winds) near the end of their lives.
Hot stars emit ionizing photons with associated photoheating that has a significant dynamical effect on their surroundings.
3D simulations show that \hii{} regions around runaway O stars drive expanding conical shells and leave underdense wakes in the medium they pass through.
For late O stars this feedback to the interstellar medium is more important than that from stellar winds.
Late in life, O stars evolve to cool red supergiants more rapidly than their environment can react, producing transient circumstellar structures such as double bow shocks.
This provides an explanation for the bow shock and linear bar-shaped structure observed around Betelgeuse.
\end{abstract}

\section{Circumstellar medium around hot stars}  
Massive stars emit strong winds and ionizing radiation during their main sequence phase, both of which strongly affect their surroundings \citep*[e.g.][]{FreHenYor03}.
The O star $\zeta$ Oph provides a dramatic example of this because of its proximity to Earth.
It is a stellar exile, moving supersonically through the interstellar medium (ISM) with velocity  $v_\star\approx26.5\,\kms$, so its stellar wind generates a bow shock when it meets the ISM \citep*[][and references therein]{GvaLanMac12}.
The separation between the star and bow shock in the upstream direction, known as the standoff distance, is $R_\mr{SO}\approx0.16$\,pc.
$\zeta$ Oph emits $3.6\times10^{47}$ ionizing photons per second, generating a large, almost spherical, photoionized \hii{} region with a radius $R_\mr{St}\approx10$\,pc.
This is about 50 times larger than $R_\mr{SO}$, so we see immediately that an O star's region of influence extends far beyond the astropause.
If we assume that the \hii{} region has only a weak effect on the ISM, we can use the relative sizes of the \hii{} region and bow shock to constrain the mass-loss rate of $\zeta$ Oph to be $\dot{M}\approx 2.2\times10^{-8}\,\msunperyr$, and the ISM nucleon number density to be $n\approx3.6\,\pcmc$ \citep{GvaLanMac12}.

\begin{figure}[t]
\vspace*{2mm}
\begin{center}
\includegraphics[width=\textwidth]{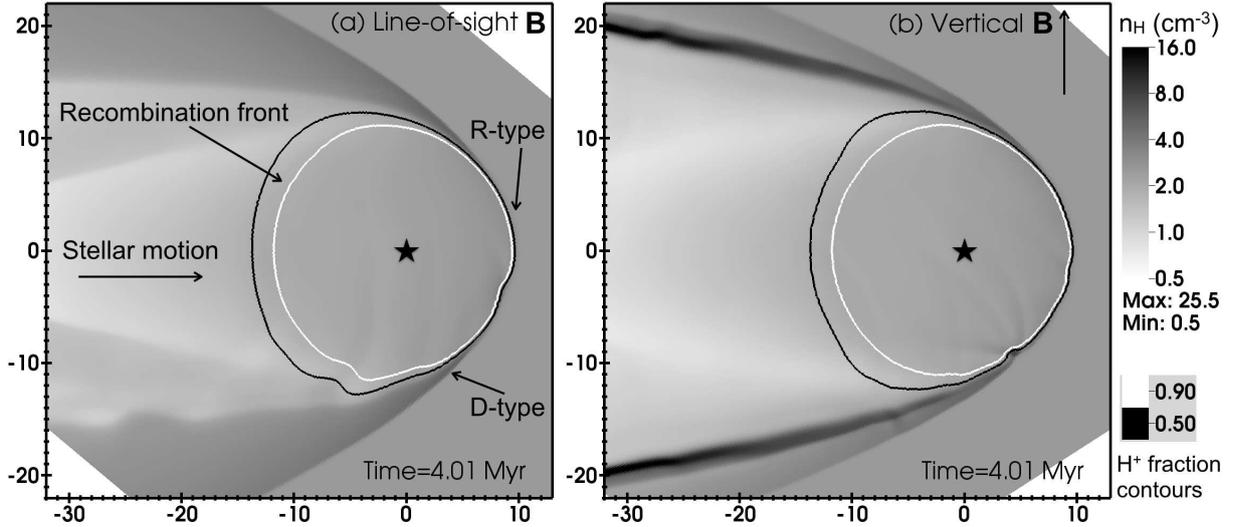}
\end{center}
\caption{
  Two perpendicular slices through a 3D R-MHD simulation of the \hii{} region produced by an O star moving through a uniform, magnetised ISM for a star that emits no stellar wind.
  The star is moving from left to right in the image plane.
  Gas number density (greyscale, in H atoms per cm$^{3}$), and ionization fraction of H  (white and black contours, indicated) are plotted.
  In panel (a) the ISM magnetic field is oriented along the line-of-sight  (perpendicular to the image plane), whereas in panel (b) it is in the image plane and vertical.
  \label{fig:zoph}
  }
\end{figure}

The bow shock of $\zeta$ Oph is much smaller than (and has negligible effect on) its \hii{} region, so we can model the two separately.
In \citet*{MacLanGva13} we have presented 3D radiation-magnetohydrodynamics (MHD) simulations of the \hii{} region around an exiled massive star moving through a uniform, magnetised medium.
The simulations use a uniform grid, with a finite-volume integration scheme for ideal MHD, coupled to a raytracer and a solver for the ionization rate equation of H.
The code calculates the non-equilibrium ionization state of H, with associated photoheating (and radiative cooling from H, He, and metals), coupled to the gas dynamics driven by the pressure difference between warm ionized gas ($T\approx10^4$\,K) and cooler neutral gas ($T\approx10^3$\,K).

Results from one simulation are shown in Fig.~(\ref{fig:zoph}), for a star moving with $v_\star=26.5\,\kms$ through a uniform ISM with H number density $n_\mr{H}=2.5\,\pcmc$.
The ISM magnetic field is initially uniform with strength $B=7\,\mu$G and oriented perpendicular to the star's direction of motion.
The ionization front (I-front) is shown by the contours of H ionization fraction.
In both panels we see the main features of the flow:
\begin{enumerate}
\item The upstream I-front is R-type \citep[rarefied;][]{Kah54}, too fast to sustain shocks, and characterised by weak density and velocity jumps, but large temperature and pressure gradients.
\item The downstream edge of the \hii{} region is a recombination front, whose width is determined by the recombination rate and the velocity of the gas through the front.
  Gas density, velocity, pressure and temperature change smoothly and gradually through the front.
\item The lateral edges of the \hii{} region have decreasing normal flow velocity through the I-front as the angle with respect to the direction of motion increases, so the I-front becomes D-type (dense) and drives a shocked shell ahead of it \citep{RagNorCanEA97}.
  The shell expands outwards in a cone-shape and leaves an underdense wake behind the star.
\end{enumerate}
Fig.~(\ref{fig:zoph}) also shows differences between the slices perpendicular (panel a) and parallel (panel b) to the magnetic field.
Panel (a) shows the shock has weaker compression and a wider opening angle than panel (b), because the anisotropic magnetic tension and fast magnetosonic speed both have a maximum in this plane.
In panel (b) the oblique shock separates the different MHD wavespeeds and so multiple  layers are formed in the expanding shell.
Shocks in this direction allow much stronger compression because the expansion is almost aligned with the magnetic field.

This shocked shell adds kinetic energy to the ISM at about the same rate as the mechanical luminosity of $\zeta$ Oph's wind, and adds momentum at more than $100\times$ the rate of the wind.
It also creates an underdense ``tunnel'' in the ISM that it passes through.
If the runaway star was ejected from a cluster of massive stars, this provides an underdense channel  from the star cluster to the Galactic halo, through which hot gas from the star cluster's supernovae can escape.

\section{Circumstellar medium around cool stars}
\begin{figure}[t]
\vspace*{2mm}
\begin{center}
\includegraphics[width=\textwidth]{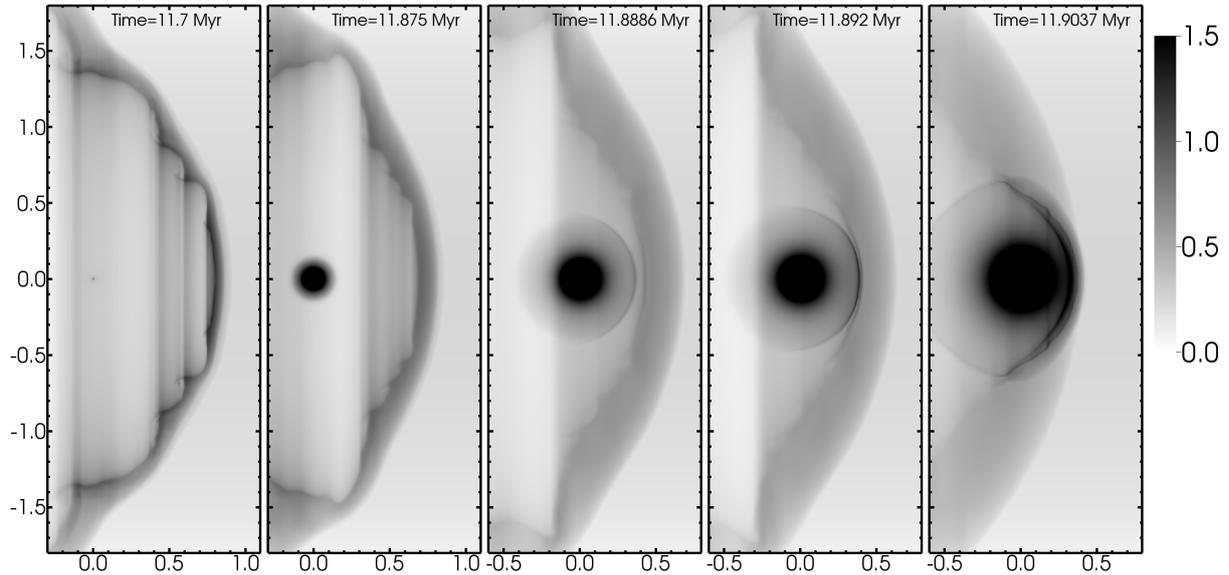}
\end{center}
\caption{
  2D simulations of the circumstellar medium at different times (indicated) in a runaway star's evolution from blue supergiant (left-most panel) to red supergiant to the pre-supernova stage (right-most panel).
  The $x$-axis is parallel to the direction of motion (from left to right), the $y$-axis is the radial direction (with rotational symmetry around $y=0$), and both axes have units of parsecs with the star at the origin.
  H column density is plotted on a linear greyscale in units of $10^{19}\,\pcms$.
  The strengthening red supergiant wind expands into the relic bow shock from the blue supergiant phase, creating a short-lived double bow shock.
  \label{fig:aori}
  }
\end{figure}

At the end of core hydrogen burning, most massive stars expand rapidly to become red supergiants on a timescale of $\approx0.01-0.02\,$Myr.
They cease to emit ionizing photons and their escape velocity decreases  by $>10\times$, so their wind becomes slower, denser, and neutral.
A new bow shock forms around the expanding slow wind within the relic bow shock from the fast wind \citep{MacMohNeiEA12}, because the star's evolution is more rapid than the dynamical timescale of the bow shock ($R_\mr{SO}/v_\star\sim0.01-0.1\,$Myr).
 Hydrodynamic simulations of this evolution are shown in Fig.~(\ref{fig:aori}), where the left-most panel shows the bow shock from the star's blue supergiant phase (with a wind velocity $v_\mr{w}\approx400\,\kms$), and subsequent panels show its evolution once the star has become a red supergiant ($v_\mr{w}\approx15\,\kms$).

This model may be able to explain the unusual circumstellar medium around the red supergiant Betelgeuse, another nearby stellar exile with a space velocity $v_\star\approx30\,\kms$ \citep{DecCoxRoyEA12}.
It has a bow shock at radius from the star $r\approx0.35$\,pc and a linear bar-shaped structure at $r\approx0.4$\,pc \citep{NorBurCaoEA97, DecCoxRoyEA12}.
The bow shock is about 40 times less massive than predicted, and so it must be very young and not yet in a steady state \citep*{MohMacLan12}.
The bar could be a chance encounter with the edge of an interstellar cloud \citep{DecCoxRoyEA12}, but its orientation perpendicular to Betelgeuse's direction of motion motivated us to ask could it be a structure generated by the star?
Our simulations show that stellar evolution can indeed leave multiple bow shocks upstream from the star.
Our model can explain the low mass and apparent youth of the bow shock and the presence of the linear bar further upstream.

\conclusions  
We have studied hydrodynamics of photoionized H\,\textsc{ii} regions around massive stars, finding that the dynamical effects of the photoheating are in some ways more important that that of the wind.
Our models for bow shocks around a blue supergiant evolving to a red supergiant show that stellar evolution modifies the stellar wind faster than the bow shock can react, leading to a short time in which there are two bow shocks around the star.
In future work we will simultaneously simulate the H\,\textsc{ii} region and bow shock around a runaway star moving through a magnetised ISM.
This will allow quantitative comparison with observations of the circumstellar medium of massive stars.

\begin{acknowledgements}
  JM acknowledges funding from a fellowship from the Alexander von Humboldt Foundation and from the Deutsche Forschungsgemeinschaft priority program 1573, ``Physics of the Interstellar Medium''.
  SM gratefully acknowledges the receipt of research funding from the National Research Foundation (NRF) of South Africa.
  HRN acknowledges funding from a NSF grant (AST-0807664).
  The authors acknowledge the John von Neumann Institute for Computing for a grant of computing time on the JUROPA supercomputer at J\"ulich Supercomputing Centre.
\end{acknowledgements}

\bibliographystyle{copernicus}
\bibliography{../../../../../documentation_misc/bibtex/refs}


%
%
%










\end{document}